\documentclass[aps,prd,twocolumn,showpacs,nofootinbib,amsmath,amssymb,amsfonts,showkeys]
{revtex4}

\usepackage{graphicx}
\usepackage{xcolor}
\usepackage{diagbox}

\begin{document}
	
	\title{Global signal in the redshifted hydrogen 21-cm line from the Dark Ages and \\ Cosmic Dawn: Dependence on the nature of dark matter and modeling of first light}
	
	% The list of authors, and the short list which is used in the headers.
	% If you need two or more lines of authors, add an extra line using \newauthor
	\author{B. Novosyadlyj$^{1,2}$, Yu. Kulinich$^2$, D. Koval$^2$}
	\affiliation{$^1$International Center of Future Science and College of Physics of Jilin University, \\  
		2699 Qianjin Str., 130012, Changchun, P.R.China, \\
		$^2$Astronomical Observatory of Ivan Franko National University of Lviv, \\
		8 Kyrylo and Methodij Str., 79005, Lviv, Ukraine }

	\date{\today}
	
	% Abstract of the paper
	\begin{abstract}
		We estimate the global signal in the redshifted hyperfine structure line 21 cm of hydrogen atoms formed during the Dark Ages and Cosmic Dawn epochs. The evolution of the brightness temperature in this line was computed to study its dependence on the physical conditions in the intergalactic medium. We show that the profile of this line crucially depends on the temperature and ionization of baryonic matter as well as the spectral energy distribution of radiation from the first sources. The cosmological models with the self-annihilating and decaying dark matter with allowable parameters by current observational data, as well as the model of the first light which is consistent with the observational data on reionization were considered. The results show that the Dark Ages part of profile is very sensitive to the parameters of self-annihilating and decaying dark matter particles, while the Cosmic Dawn part of profile is very sensitive also to the spectral energy distribution of radiation from the first sources. It was concluded that only compatible observations of the redshifted 21 cm line in the decameter and meter wavelength range, formed during the Dark Ages and Cosmic Dawn, will make it possible to constrain the parameters of dark matter models and astrophysical models of the first sources based on the radiotomography of the young Universe.
	\end{abstract}
	\maketitle

	%%%%%%%%%%%%%%%%%%%%%%%%%%%%%%%%%%%%%%%%%%%%%%%%%%

	\section{Introduction}
	
	The epoch of the Universe's history after cosmological recombination, when the first stars and galaxies began to form, is terra incognito from the observational point of view. Its discovery is expected in the coming decades, when a number of planed groundbased, space and lunar sensitive telescopes  for meter and decameter wavelength will be implemented \cite{Bale2023,Burns2021,Goel2022,Bentum2020,Shkuratov2019,Sokolowski2015,deLera2022,Nhan2019,Monsalve2023,Philip2019,Burns2011,SKA}. The matter is that the hyperfine structure line of hydrogen 21 cm is the main channel of information about state of baryonic gas in that period but it is redshifted by cosmological expansion up to meter - decameter wavelength range complicated for extragalactic observations. 
	
	A lot of researchers, starting from S. Wouthuysen \cite{Wouthuysen1952} and G.B. Field \cite{Field1958,Field1959} in 50th years of the last century up to now (see reviews \cite{Barkana2001,Fan2006,Furlanetto2006,Bromm2011,Pritchard2012,Natarajan2014,Shimabukuro2022,Minoda2023} studied the different aspects of the  manifestations of this line at the cosmological space-time scales and the possibility of its instrumental detection. They show that the standard thermal model of the evolution of the Universe predicts the existence of three spectral features: the Dark Ages absorption line at decameter wavelenghts 10-40 m ($z\sim50-200$), the Cosmic Dawn absorption line at meter wavelengths 2-7 m ($z\sim10-30$), and the emission line in the Epoch of Reionization ($\sim6-10$). The second absorption line is caused by the Wouthuysen-Field effect and is determined by the spectral energy distribution (SED) of the first sources radiation (the first light). The first announcement about the detection of this line in the Experiment to Detect the Global Epoch of Reionization Signature experiment (EDGES)\cite{Bowman2018} indicated an unexpectedly deep absorption line at 78 MHz: $\sim$2 times deeper than the maximal one predicted in the framework of standard cosmology \cite{Cohen2017,Reis2021}. These results, however, are criticized by other experts. For example the authors of \cite{Hills2018,Singh2019,Sims2020,Bevins2021} suggest that this may be due to the difficulty of completely removing the influence of the galactic foreground radiation and other interferences, the intensity of which exceeds the signal from the Cosmic Dawn. Recently, the team of  Shaped Antenna measurement of the background RAdio Spectrum 3 (SARAS3) experiment announced the results of similar measurements  \cite{Singh2022}: they reject the best-fitting profile obtained by EDGES, with 95.3\% confidence. 
	
	It is important to know the sensitivity of these spectral features in the redshifted 21cm line to variations of the parameters of cosmological models and the first light. A number of articles are devoted to this aspect, in particular \cite{Pritchard2010,Cohen2017,Barkana2018,Ewall-Wice2018,Reis2021,Halder2022,Novosyadlyj2023,Facchinetti2024,Novosyadlyj2024}. It was shown that the positions and depths/heights of the absorption/emission lines formed during Cosmic Dawn and Reionization epochs practically completely fill in the brightness temperature - redshift space in rectangle $T_\mathrm{br}^{min}\approx-200$ mK, $T_\mathrm{br}^{max}\approx20$ mK, $z_\mathrm{max}\approx20$, $z_\mathrm{min}\approx8$ for reasonable values of the number density of $Ly_\alpha$, UV- and X-quanta \cite{Cohen2017,Reis2021}. The standard $\Lambda$CDM model with Planck parameters predicts a value of the brightness temperature for the signal from the Dark Ages epoch at the center of absorption line $T_\mathrm{br}\approx-35$ mK at $z\approx87$ \cite{Novosyadlyj2023}. The depth of the line is moderately sensitive to $\Omega_b$ and $H_0$, weakly sensitive to $\Omega_\mathrm{dm}$ and insensitive to other parameters of the standard $\Lambda$CDM model \cite{Novosyadlyj2023,Novosyadlyj2024}. But it is very sensitive to additional heating/cooling and ionization of baryonic component. A model with additional cooling leads to a deeper depth of absorption line. Heating and ionization take place, in particular, in models with decaying or self-annihilating dark matter \cite{Valdes2013,Evoli2014,DAmico2018,Clark2018,Mitridate2018,Hiroshima2021,Takahashi2021,Liu2023a,Liu2023b,Sun2023,Novosyadlyj2023,Facchinetti2024,Novosyadlyj2024}, and cooling in models with dark matter whose particles interact with baryonic matter through the weak Coulomb-like force \cite{Munoz2015,Ali2015,Essig2017,Barkana2018}. Heating and ionization of baryonic matter lead to shallowing of the redshifted 21 cm absorption line or turning it into emission one.  
	
	The papers cited above are mainly focused on the investigation of 21 cm global signal from the Dark Ages or Cosmic Dawn/Reionization epochs separately. Since the Wouthuysen-Field effect is crucial for the formation of the spectral features associated with the redshifted 21 cm line in Cosmic Dawn and Reionization epochs the radiation of the first sources of the $Ly_\mathrm{\alpha}/Ly_\mathrm{c}$ quanta and X-ray define the frequency and amplitude of spectral feature at 50-100 MHz \cite{Cohen2017,Reis2021}. Significant dispersion of peak frequencies and amplitudes, and spatial ingomogeneities stimulate us to thought about informativeness of global signal from the Cosmic Dawn and Reionization epochs. The global signal from the Dark Ages at 10-50 MHz is moderately sensitive to cosmological parameters of the standard CDM model and strongly sensitive to the additional mechanizm of heating, ionization and cooling caused by the nature of dark matter particles, existing of primordial black holes or magnetic fields. Wide spectral tomography 10-100 MHz may strong distinguish the  includes of different mechanizm. In this paper we move to the analysis of such a possibility.
	
	In this paper we estimate the dependence of global signal in the redshifted 21 cm line formed in the Dark Ages and Cosmic Dawn epochs on the model parameters of the decaying and self-annihilating dark matter. We approximate the evolution model of the energy density  of the first light  by black-body radiation with evolving temperature. Its parameters were constrained by the observational data on reionization following from CMB and distant quasars and galaxies spectra measurements.
	
	The outline of the paper is as follows. In Section 2 we describe the models of the dark energy decaying, self-annihilating and the first light and compute the evolution of temperature and ionization state of the baryonic gas for Dark Ages and Cosmic Dawn up to the completion of Reionization. In Section 3 we analyse the sensitivity of the positions and depths of the Dark Ages and Cosmic Dawn    absorption lines to variations of the  model parameters of dark matter and the first light. The results are discussed and summarised in Section 4.
	
	\section{State of atomic hydrogen from cosmological recombination to the completion of Reionization}
	
	\begin{figure}[htb!]
		\includegraphics[width=0.5\textwidth]{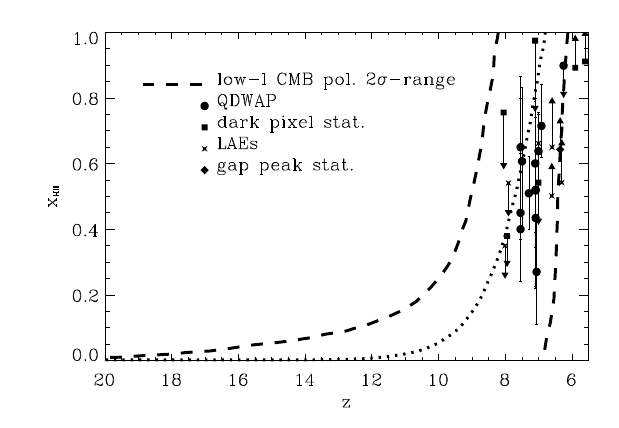}
		\caption{Fraction of ionized hydrogen at Cosmic Dawn and Reionization epochs from cosmological and astrophysical observational data. The dashed lines show the 2$\sigma$-range given by Planck Collaboration \citep{Planck2020}, the dotted line is the median value \cite{Glazer2018}, the circles show the estimations derived from the damping wing absorption profiles in the spectra of quasars (QDWAP)  \cite{Schroeder2013,Greig2017,Mortlock2011,Davies2018,Bouvens2015,Banados2018,Greig2022}, the squares -- from the dark pixel statistics \cite{McGreer2015}, 4-fold stars -- from the redshift-dependent prevalence of $Ly_\alpha$ emitters (LAEs) \cite{Schenker2014,Mason2018,Mason2019a,Ouchi2010}, the diamonds -- from the gap/peak statistics \cite{Gallerani2008}.} \label{rei}
	\end{figure}
	For the computation of the global signal in the hydrogen 21 cm line from the Dark Ages and Cosmic Dawn epochs, it is necessary to precompute the ionization and thermal history of the gas under some assumption about first light model. Such history depends on the parameters of the cosmological model, the ionization mechanisms, heating and cooling of the baryonic gas. In this section, we will consider the ionization and thermal history of the gas in the standard cosmological model with stable cold dark matter particles and in the models with decaying and self-annihilating ones. The standard cosmological model to be $\Lambda$CDM with cosmological parameters following from \cite{Planck2020,Planck2020a}: $H_0=67.36\pm0.54$ km/s$\cdot$Mpc (Hubble constant), $\Omega_b=0.0493\pm0.00112$ (density parameter\footnote{Density parameter is the mean density of the component in the current epoch in units of critical: $\Omega_i\equiv\rho^0_i/\rho^0_\mathrm{cr}$ where $\rho^0_\mathrm{cr}\equiv 3H_0^2/8\pi G$.} of baryonic matter), $\Omega_\mathrm{dm}=0.266\pm0.0084$ (density parameter of dark matter), $\Omega_r=2.49\cdot10^{-5}$ (density parameter of relativistic component (CMB and neutrinos)), $\Omega_\mathrm{\Lambda}=0.6847\pm0.0073$ and $\Omega_K=0$ (dimensionless\footnote{$\Omega_\mathrm{\Lambda}\equiv c^2\Lambda/3H^2_0$, $\Omega_\mathrm{K}\equiv -c^2K/H^2_0$} cosmological constant $\Lambda$ and curvature of 3-space $K$, respectively). We assume the primordial chemical composition with mass fraction of helium $Y_p=0.2446$ \citep{Peimbert2016}, following from the cosmological nucleosynthesis of standard Big Bang model.
	
	In the homogeneous expanding Universe, the kinetic temperature of the gas and the ionization state of atoms at any redshift we compute by integrating a system of differential equations of energy balance and ionization-recombination kinetics of atoms.  The first molecules in the intergalactic medium of early Universe make up a small fraction of all particles (see \cite{Puy1993,Galli1998,Vonlanthen2009,Novosyadlyj2017,Kulinich2020,Novosyadlyj2020a,Novosyadlyj2022} and citation therein) and are not discussed in this work. In the standard $\Lambda$CDM  model, the main mechanizm of heating of the gas during Dark Ages is the Compton scattering of the CMB radiation on free electrons, and main mechanizm of cooling is the adiabatic cooling via cosmological expansion. All other mechanisms that occur in astrophysical plasma under the conditions of the Dark Ages are weaker by 3 or more orders of magnitude \cite{Novosyadlyj2024}. The temperature of the plasma and the ionization of hydrogen and helium during cosmological recombination and in the epoch of the Dark Ages up to $z=200$ are computed based on an effective 3-level atomic model implemented in the publicly available program RecFast\footnote{http://www.astro.ubc.ca/people/scott/recfast.html} \cite{Seager1999,Seager2000}. During the Cosmic Dawn epoch, the heating by radiation of the first sources via Compton scattering and photoinization as well as cooling via photorecombination anf free-free transition became important too \cite{Novosyadlyj2023}. For $z<200$, we used the following system of equations  
	\begin{eqnarray}
		&&\hskip-0.5cm -\frac{3}{2}n_\mathrm{tot}k_\mathrm{B}(1+z)H\frac{dT_\mathrm{b}}{dz} = \Gamma_\mathrm{C_\mathrm{CMB}}-\Lambda_\mathrm{ad}+\Gamma_\mathrm{C_\mathrm{fl}}+\Gamma_\mathrm{phi}\nonumber \\
		&&\hskip2.5cm-\Lambda_\mathrm{phr}-\Lambda_\mathrm{ff}-\Lambda_\mathrm{etc} + \Gamma_\mathrm{nSM},\label{Tb}\\
		&&\hskip-0.5cm -(1+z)H\frac{dx_\mathrm{HII}}{dz}=R_\mathrm{HI}x_\mathrm{HI}+C^i_\mathrm{HI} n_i x_\mathrm{HI}-  \nonumber \\
		&&\hskip1.0cm-\alpha_\mathrm{HII}x_\mathrm{HII}x_\mathrm{e} n_\mathrm{H}-(1+z)H\frac{dx_\mathrm{HI}}{dz}|_\mathrm{nSM}, \label{kes}\\
		&&\hskip-0.5cm -(1+z)H\frac{dx_\mathrm{HeII}}{dz}=R_\mathrm{HeI}x_\mathrm{HeI}+C^i_\mathrm{HeI} n_i x_\mathrm{HeI}- \nonumber \\
		&&\hskip0.5cm-\alpha_\mathrm{HeII}x_\mathrm{HeII}x_\mathrm{e} n_\mathrm{H}-(1+z)H\frac{dx_\mathrm{HeI}}{dz}|_\mathrm{nSM}, \label{kes2}\\
		&&\hskip-0.5cm H=H_0\sqrt{\Omega_m(1+z)^3+\Omega_K(1+z)^2+\Omega_\mathrm{\Lambda}}, \label{H}
	\end{eqnarray}
	where $T_\mathrm{b}$ is the temperature of the baryonic gas, $n_\mathrm{tot}\equiv n_\mathrm{HI}+n_\mathrm{HII}+n_\mathrm{e}+n_\mathrm{HeI}+n_\mathrm{HeII}=n^0_\mathrm{tot}(1+z)^3$ is the number density of particles of all types, $n_\mathrm{H}\equiv n_\mathrm{HI}+n_\mathrm{HII}=n^0_\mathrm{H}(1+z)^3$ is the number density of hydrogen, $k_\mathrm{B}$ is the Boltzmann constant, $\Gamma_\mathrm{C_\mathrm{CMB}}$ is the heating function due to Compton scattering of CMB quanta on free electrons \cite{Seager1999,Seager2000},    
	$\Lambda_\mathrm{ad}$ is the adiabatic cooling function due to the expansion of the Universe, $\Gamma_\mathrm{C_\mathrm{fl}}$ is the heating function due to Compton scattering of the quanta of the first light on free electrons, $\Gamma_\mathrm{phi}$ is heating function by the photoionization,  $\Lambda_\mathrm{phr}$ is the cooling function due to photorecombination, $\Lambda_\mathrm{ff}$ is the cooling function due to free-free transitions of electrons, $\Lambda_\mathrm{etc}$ collects all other cooling function which are presented in the Appendix of \cite{Novosyadlyj2023} and give negligibly contribution into energy balance in our cases. The equations (\ref{Tb})-(\ref{kes2}) can be found in other papers about physical conditions in these epochs, especially see in \cite{Naoz2005,Chluba2010,Liu2018,Liu2019,Flitter2024}. Eq. (\ref{H}) is well known Friedmann equation.
	
	We integrated the system of equations (\ref{Tb})-(\ref{H}) and computed spin and brightness temperatures using our code H21cm.f described in \cite{Novosyadlyj2023}. The analytical approximations of the heating/cooling functions used here are collected there in Appendix A and their contributions to the energy balance equation are shown in Figure 4. X-ray and $Ly_\alpha$ heating are not included. So, our results are related to the models with no domination of these heating mechanisms in the Cosmic Dawn epoch. The heating functions in non-standard models $\Gamma_\mathrm{nSM}$ are presented below. 
	
	In the equations of ionization-recombination kinetics, $R_\mathrm{HI},\,R_\mathrm{HeI}$ are the photoionization rates of hydrogen and helium atoms, $C^i_\mathrm{HI},\,C^i_\mathrm{HeI}$ are the collisional ionization coefficients of the $i$-th type of particles, $\alpha_\mathrm{HII},\,\alpha_\mathrm{HeII}$ are the recombination coefficients of hydrogen and helium ions, $\Omega_m=\Omega_b+\Omega_\mathrm{dm}$, $H(z)$ is the expansion rate of the Universe. The last terms there are the rates of additional ionization in the non-standard models of dark matter.
	
	\subsection{Model of the first light}
	
	The most scenarios of the formation of the first light sources assume that they were stars of the first generation, Population III. Let's assume that the background radiation consist of the CMB and light of the first stars, and its total spectral energy density (SED) can be presented as 
	\begin{equation}
		J_\mathrm{fl}(\nu)=\frac{4\pi}{c}\left[B(\nu;T_\mathrm{CMB})+\sum_i\mathcal{\beta}^{(i)}_\mathrm{fl}B(\nu;T^{(i)}_\mathrm{fl})\right], \label{jnu}
	\end{equation} 
	where $B(\nu,T)$ is the Planck function, the coefficient $\beta^{(i)}_\mathrm{fl}$ is dilution coefficient, $T^{(i)}_\mathrm{fl}$ is an effective surface temperature of the first stars. Their total SED can be approximated by combination of several Planck function with different temperature  ($i=1,\,2\,...$ marks each from them). The evolution of SED can be described though z-dependence of $T^{(i)}_\mathrm{fl}$, or $\beta^{(i)}_\mathrm{fl}$, or both. Here, as in \cite{Novosyadlyj2023}, we model it by tanh-dependence of temperature $T^{(i)}_\mathrm{fl}$ on redshift:
	\begin{equation}
		T_\mathrm{fl}^{(i)}=T_*^{(i)}\tanh{\left[a_\mathrm{fl}^{(i)}\left(\frac{1+z_\mathrm{fl}^{(i)}}{1+z}\right)^{b_\mathrm{fl}^{(i)}}\right]}.\label{Tfl}
	\end{equation} 
	All parameters, $\beta^{(i)}_\mathrm{fl}$, $a^{(i)}_\mathrm{fl}$, $b^{(i)}_\mathrm{fl}$ and $z^{(i)}_\mathrm{fl}$ are fitted to obtain the $x_\mathrm{HII}(z)$ matching the observational data for it (Fig. \ref{rei}). The parameter $T_\mathrm{*}^{(i)}$ is set manually. In this paper, we use three models of the first light with different SEDs, which provide the evolution of $x_\mathrm{HII}(z)$ shown by dashed and dotted lines in Fig. \ref{rei}. They correspond to the high-z and low-z contours of the reionization range established by the Plank team \cite{Planck2020a}, and intermediate z-track which corresponds to the median dependence $x_\mathrm{HII}(z)$ by \cite{Glazer2018}. The values of the parameters of such models of the first light and their designations are given in the table I.
	\begin{table}[h] 
		\begin{center}
			\caption{Parameters of models of the first light}  
			\begin{tabular} {c|ccccc}
				\hline
				\hline
				\noalign{\smallskip}
				Model&$T_*$ (K)&$\beta_\mathrm{fl}$&$z_\mathrm{fl}$&$a_\mathrm{fl}$&$b_\mathrm{fl}$ \\
				\noalign{\smallskip} 
				\hline
				\noalign{\smallskip} 
				fl2a&5000&$2.7\cdot10^{-10}$&0.0&0.5&6.0\\
				&20000&$1.0\cdot10^{-19}$&4.2&3.0&5.8\\
				\noalign{\smallskip}    
				\hline 
				fl2b &5000&$1.0\cdot10^{-10}$&5.8&6.0&2.4\\ 
				&20000&$4.5\cdot10^{-19}$&3.8&6.0&3.8\\
				\noalign{\smallskip}    
				\hline
				fl2c &5000&$1.0\cdot10^{-11}$&4.6&6.0&2.5\\ 
				&20000&$8.0\cdot10^{-19}$&4.8&2.5&5.0\\
				\noalign{\smallskip}    
				\hline  
				\hline
			\end{tabular}
		\end{center}
		\label{pmfl2}
	\end{table}
	
	\noindent The first light model fl2a provide early, fl2b median and fl2c late reionizations shown in Fig. \ref{rei}. The SEDs of radiation (\ref{jnu}) for these models of the first light at the Cosmic Dawn and Reionization epochs are shown in Fig. 10 in \cite{Novosyadlyj2023}.

	\subsection{Model with self-annihilating dark matter particles}
	
	\begin{figure*}[htb] 
		\includegraphics[width=0.47\textwidth]{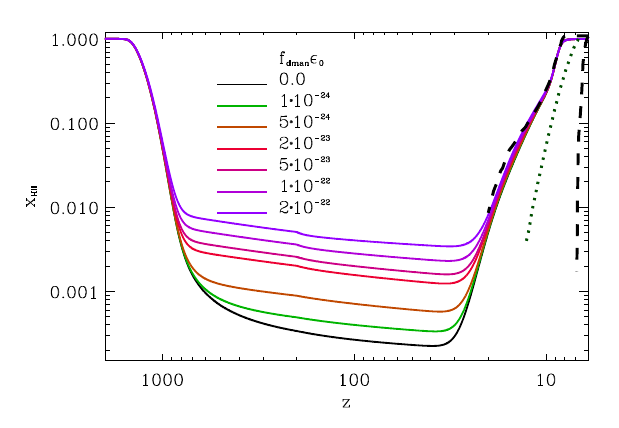}
		\includegraphics[width=0.47\textwidth]{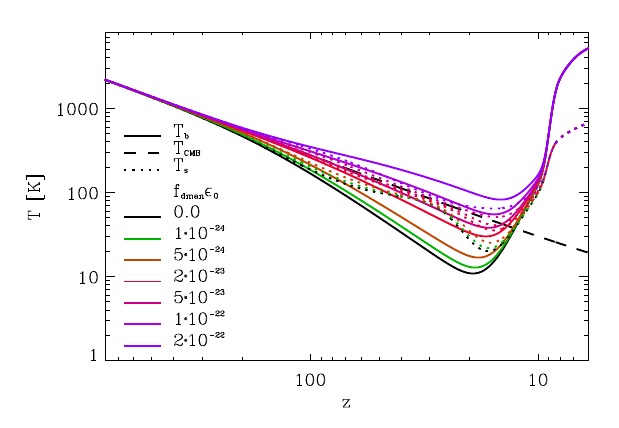}
		\includegraphics[width=0.47\textwidth]{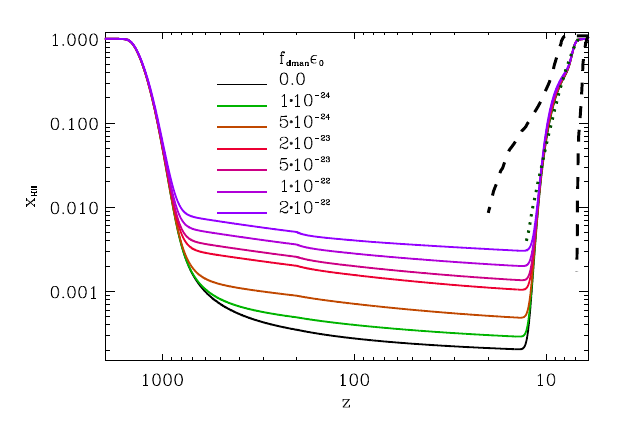} 
		\includegraphics[width=0.47\textwidth]{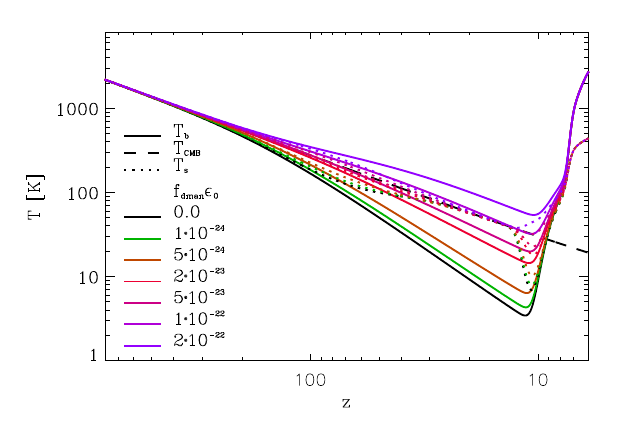}
		\includegraphics[width=0.47\textwidth]{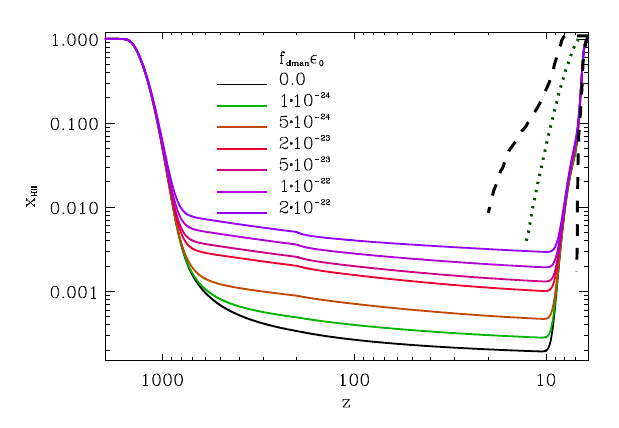} 
		\includegraphics[width=0.47\textwidth]{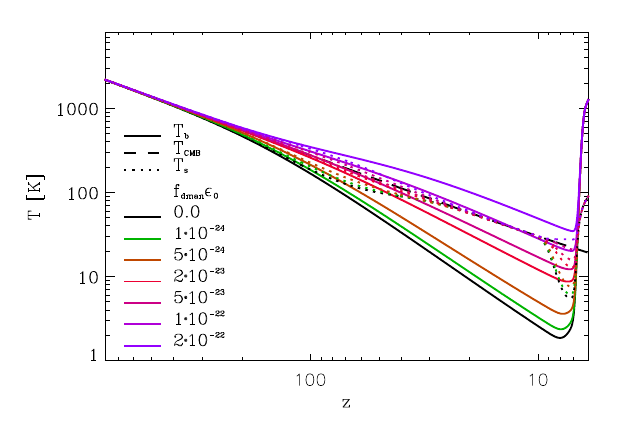}
		\caption{The dependences of the ionized hydrogen fraction (left column) and kinetic and excitation temperatures (right column) on redshift in the models with self-annihilating dark matter with different values of the consolidated parameter $f_\mathrm{dman}\epsilon_0$, which increases for lines from bottom to top, and three models of the first light flIIa, flIIb and flIIc (figures from top to down). The dashed and dotted lines in the left column figures are the same as in Fig. \ref{rei}.}
		\label{xT_dman}
	\end{figure*}
	
	Standard $\Lambda$CDM model has only one parameter for dark matter which corresponds its average mass density in the Universe. Other parameters of dark matter follow from assumptions about its nature. Upper limits were established for them based on the results of astrophysicalobservations and laboratory experiments. Here we  search the values of the parameters of self-annihilating dark matter particles, at which the change in the ionization and thermal history of baryonic matter in the Dark Ages and Cosmic Dawn epochs would manifest itself in the characteristics of the hydrogen 21 cm line. For this, we will use a simple model of self-annihilating dark matter, which is proposed by Jens Chluba \cite{Chluba2010}. It is assumed that the dark matter particles with mass $m_\mathrm{dm}$ self-annihilate with an effective cross-section $\langle\sigma v\rangle$, averaged over their velocities. The energy that goes to heating of baryonic matter per unit volume per unit time is equal to
	\begin{equation}
		\Gamma_\mathrm{dman}=1.6\cdot10^{-12}f_\mathrm{dman}g_\mathrm{h} \epsilon_0n_\mathrm{H}(1+z)^3 \quad \frac{\mbox{\rm erg}}{\mbox{\rm cm$^3$s}}, \label{Gan}
	\end{equation}
	where
	\begin{equation}
		\epsilon_0=4.26\cdot10^{-28}\left[\frac{100\mathrm{GeV}}{m_\mathrm{dm}}\right]\left[\frac{\Omega_\mathrm{dm}h^2}{0.12}\right]^2\left[\frac{\langle\sigma v\rangle}{10^{-29}\mbox{\rm cm$^3$/s}}\right],\nonumber
	\end{equation} 
	is a dimensionless parameter of the self-annihilation of dark matter particles, $f_\mathrm{dman}$ is the fraction of the released energy absorbed by baryonic matter, and $g_\mathrm{h}=(1 + 2x_\mathrm{HII} + f_\mathrm{He}(1 + 2x_\mathrm{HeII}))/3(1+f_\mathrm{He})$ is the fraction of that energy that goes to heating of the gas.
	
	The products of the annihilation of dark matter particles with an energy of $c^2m_\mathrm{dm}\gg20$ eV are capable of ionizing the hydrogen and helium atoms. Within the framework of the model \cite{Chluba2010}, the ionization rate can be computed as follows:
	\begin{eqnarray*}
		(1+z)H\frac{dx_\mathrm{HI}}{dz}|&=&0.0735f_\mathrm{dman}g_\mathrm{ion}^\mathrm{(HI)}\frac{\epsilon_0n_\mathrm{H}(1+z)^3}{n_\mathrm{HI}(1+f_\mathrm{He})}, \\
		(1+z)H\frac{dx_\mathrm{HeI}}{dz}&=&0.04065f_\mathrm{dman}g_\mathrm{ion}^\mathrm{(HeI)}\frac{\epsilon_0n_\mathrm{H}f_\mathrm{He}(1+z)^3}{n_\mathrm{HeI}(1+f_\mathrm{He})},
	\end{eqnarray*}
	where $g_\mathrm{ion}^\mathrm{(HI)}=(1-x_\mathrm{HII})/3$ and $g_\mathrm{ion}^\mathrm{(HeI)}=(1-x_\mathrm{HeII})/3$ are the fractions of the released energy that go to the ionization of hydrogen and helium, respectively. As we can see, $g_\mathrm{h}\propto g_\mathrm{ion}\propto 1/3$, which means that approximately one-third of the energy injected into the baryonic component goes to heating, approximately one-third to the ionization of hydrogen and helium, and the rest to the excitation of the energy levels of neutral atoms. Since the phenomenological model \cite{Chluba2010} does not propose a specific model of dark matter particles that self-annihilate, we will estimate the ionization and temperature of the baryonic gas for different values of the consolidated dimensionless parameter $f_\mathrm{dman}\epsilon_0$. 
	
	The results are shown in Fig. \ref{xT_dman}. Since the annihilation rate of dark matter particles is proportional to the square of their number density, the contribution to the ionization and thermal history of baryonic matter in the Dark Ages epoch is significant for values of the combined parameter $f_\mathrm{dman}\epsilon_0\ge10^{-24}$. In the Cosmic Dawn epoch, the ionization and heating by the first light is determining. The kinetics of cosmological recombination becomes noticeably different from the standard one in models with $f_\mathrm{dman}\epsilon_0\gtrsim10^{-22}$. In such a case the temperature of the baryonic component it turns out greater than the temperature of the CMB at $z<200$. 
	
	\subsection{Model with decaying dark matter particles}
	
	\begin{figure*}[htb!]
		\includegraphics[width=0.47\textwidth]{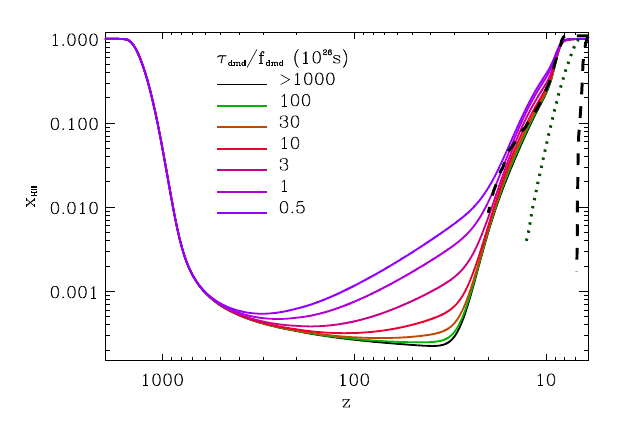}
		\includegraphics[width=0.47\textwidth]{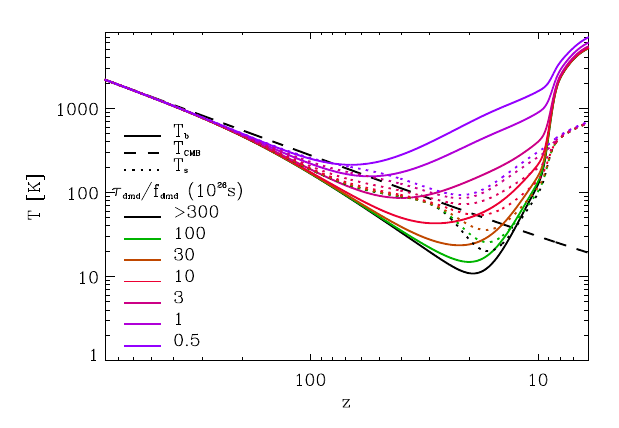}
		\includegraphics[width=0.47\textwidth]{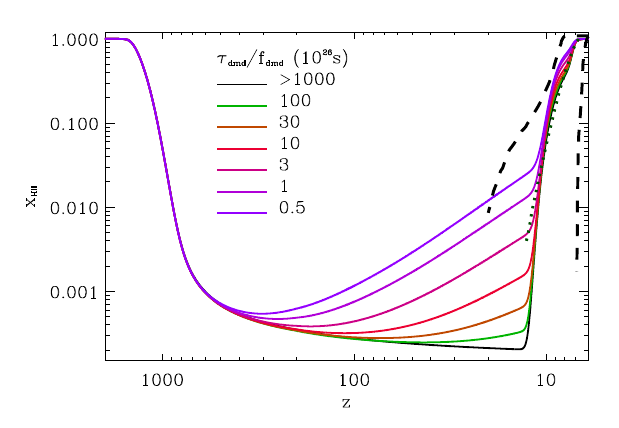}
		\includegraphics[width=0.47\textwidth]{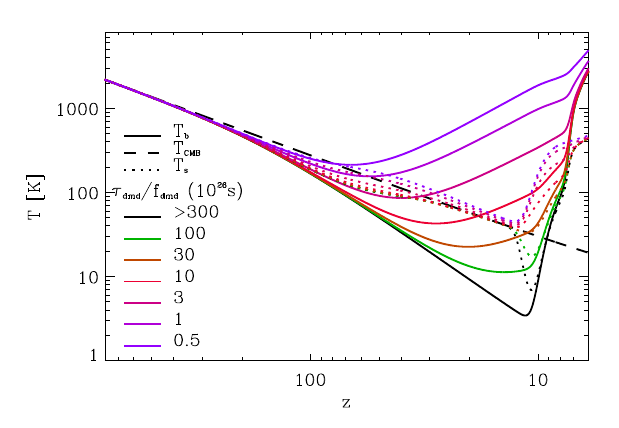}
		\includegraphics[width=0.47\textwidth]{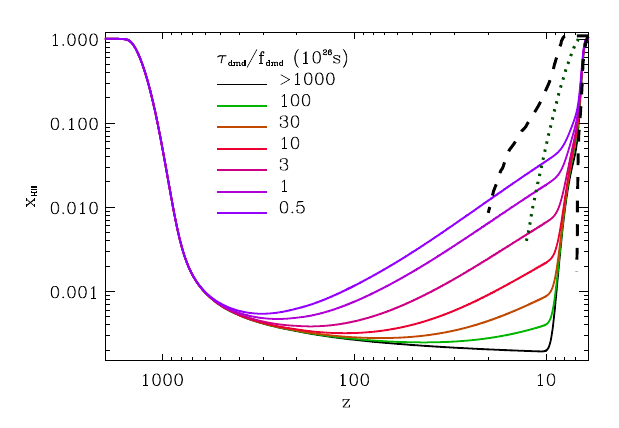}
		\includegraphics[width=0.47\textwidth]{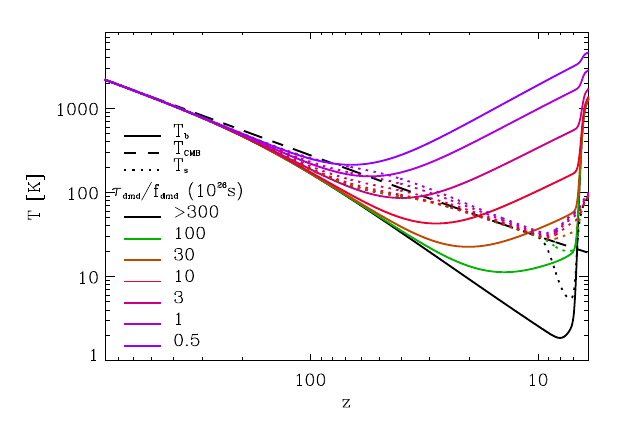}
		\caption{The dependences of the ionized hydrogen fraction (left column) and kinetic and excitation temperatures (right column) on redshift in the models with decaying dark matter with different values of the consolidated parameter $\tau_d/f_\mathrm{dmd}$, which decreases for lines from bottom to top, and three models of the first light flIIa, flIIb and flIIc (figures from top to down). The dashed and dotted lines are the same as in Fig. \ref{rei}.}
		\label{xT_dmd}
	\end{figure*}
	Dark matter particles can be decaying particles with a lifetime much larger than the age of the Universe. The key parameters of such model are the fraction of dark matter that decays, its characteristic lifetime, and the fractions of absorbed energy that go into heating, ionization, and excitation of atoms. Here, as in previous models, we will estimate the values of the decaying dark matter parameters for which the ionization and thermal history of baryonic matter in the Dark Ages epoch differs from the standard one, but does not contradict observational constraints on the optical depth of reionization. To do this, we will use the phenomenological decaying dark matter model \cite{Liu2018}, according to which the heating function can be represented as follows
	\begin{eqnarray}
		\Gamma_\mathrm{dmd}=1.69\cdot10^{-8}f_\mathrm{dmd}g_\mathrm{h}\left[\frac{\Omega_\mathrm{dm}h^2}{0.12}\right]\frac{(1+z)^3}{\tau_\mathrm{dmd}} \,\, \frac{\mbox{\rm erg}}{\mbox{\rm cm$^3$s}}, 
	\end{eqnarray} 
	where $\tau_\mathrm{dmd}$ is the decay lifetime of dark matter particles in seconds, $f_\mathrm{dmd}$ is the fraction of the released energy that is absorbed by baryonic matter, $g_\mathrm{h}$ is the fraction of absorbed energy that goes to heating of the gas. 
	
	Products of the decay of dark matter particles with energies in the keV--GeV range ionize the hydrogen and helium atoms.  The ionization rate in this case can be computed as follows \cite{Ouchi2010,Novosyadlyj2024}: 
	\begin{eqnarray*}
		(1+z)H\frac{dx_\mathrm{HI}}{dz}&=&7.35\cdot10^{-2}\frac{g_\mathrm{ion}^\mathrm{(HI)}}{g_\mathrm{h}}\frac{n_\mathrm{H}\Gamma_\mathrm{dmd} }{n_\mathrm{HI}(1+f_\mathrm{He})}, \\
		(1+z)H\frac{dx_\mathrm{HeI}}{dz}&=&4.065\cdot10^{-2}\frac{g_\mathrm{ion}^\mathrm{(HeI)}}{g_\mathrm{h}}\frac{n_\mathrm{H}f_\mathrm{He}\Gamma_\mathrm{dmd}}{n_\mathrm{HeI}(1+f_\mathrm{He})},
	\end{eqnarray*}
	where the fractions of the released energy that go to heating and ionization of hydrogen and helium, $g_\mathrm{h}$, $g_\mathrm{ion}^\mathrm{(HI)}$ and $g_\mathrm{ion}^\mathrm{(HeI)}$, are computed in the same way as for the previous case of self-annihilating dark matter. 
	
	The results of the calculations of ionization and thermal history of the gas from the epoch of cosmological recombination through the Dark Ages to complete reionization at $z=6$ for values of the consolidated parameter $\tau_\mathrm{dmd}/f_\mathrm{dmd}\in5\cdot10^{25} - 10^{28}$s are presented in Fig. \ref{xT_dmd}. For values of this parameter greater than $3\cdot10^{28}$ s, the ionization and thermal history of the gas is practically indistinguishable from the history in the standard $\Lambda$CDM. For the values less than $10^{25}$ s, the ionization of hydrogen at $z\sim20-10$ is higher than the 2$\sigma$ upper limit set in the Planck experiment \cite{Planck2020a} (dashed line in the left panel of Fig. \ref{xT_dmd}). 
	
	The evolution of the gas temperature in this model, presented in the right panel of Fig. \ref{xT_dmd}, is significantly different from the evolution in the standard $\Lambda$CDM and in the model with self-annihilating dark matter.  
	
	\section{Global signal in 21 cm line from Dark Ages and Cosmic Dawn }
	
	The signal in 21 cm line from the Dark Ages and Cosmic Dawn epochs can be a source of information about the ionization and thermal history of baryonic matter in the early Universe as well as about SED of the first sources of radiation, since the spectral features of this signal depends on the number density of neutral hydrogen, temperature of the gas and CMB radiation as well as the field of radiation from the first sources. The spin temperature $T_\mathrm{s}$, which reflects the population of the hyperfine structure levels of hydrogen, is determined by the processes of excitations and deactivations by CMB photons and collisions with electrons, protons, and neutral hydrogen atoms. In the Cosmic Dawn epoch, when a number density of $Ly_\alpha$ quanta from the first sources reaches some critical value, the Wouthuysen-Field coupling become to determine the spectral features in the 21 cm line signal. We use  a well-reasoned expression \cite{Field1958,Furlanetto2006,Pritchard2012}
	\begin{eqnarray}
		&&T_\mathrm{s}^{-1} = \frac{T_\mathrm{CMB}^{-1} + x_\mathrm{c} T_\mathrm{b}^{-1} + x_\alpha T_\mathrm{c}^{-1}}{1+x_\mathrm{c}+x_\alpha}, \label{TsCD}\\
		&&x_\mathrm{c} \equiv \frac{C_\mathrm{10}}{A_\mathrm{10}}\frac{h_\mathrm{P}\nu_\mathrm{21}}{k_\mathrm{B} T_\mathrm{CMB}}, \quad x_\alpha \equiv \frac{8\pi c^2\Delta\nu_\alpha}{9A_\mathrm{10}h_\mathrm{P}\nu_\alpha^3}S_\alpha J_\alpha,
		\nonumber
	\end{eqnarray}
	where $x_\mathrm{c}$ is the collisional coupling parameter, $h_\mathrm{P}$ is the Planck constant, $\nu_\mathrm{21}$ is the laboratory frequency of the 21 cm line, $A_\mathrm{10}$ is the Einstein spontaneous transition coefficient, $C_\mathrm{10}$ is the collisional deactivation rate by electrons, protons, and neutral hydrogen atoms, $x_\alpha$ is $Ly_\alpha$ coupling parameter, $\nu_\alpha$ is the frequency of the  $Ly_\alpha$ line, $\Delta\nu_\alpha$ is its half-width, $S_\alpha$ is scattering function of $Ly_\alpha$ quanta and $J_\alpha$ is their energy density, $T_\mathrm{c}$ is colour temperature in the line. $J_\alpha$ is computed using eqs. (\ref{jnu})-(\ref{Tfl}),  $S_\alpha$ and $T_\mathrm{c}$ are computed using the analytic approximation formulae (40)-(42) from \cite{Hirata2006}. They approximate the numerical results with an accuracy $\sim1\%$ in the range temperatures $T_\mathrm{b}\ge2$ K, $T_\mathrm{s}\ge2$ K and Gunn-Peterson optical depth $10^5\le \tau_\mathrm{GP}\le10^7$, where 
	$$\tau_\mathrm{GP}=4.67\cdot10^5\frac{\Omega_\mathrm{b} h}{\Omega_\mathrm{m}^{1/2}}(1-Y_p)x_\mathrm{\rm HI} (1+z)^{3/2},$$
	that follows from eq. (35) in \cite{Hirata2006}. The last formula shows that at the redshift of complete reionization at $z\ge6$ the accuracy of the approximation is worse, but there line disappears since $x_\mathrm{HI}\rightarrow0$.
	In Figs. \ref{xT_dman}, \ref{xT_dmd} the dotted lines in the left panels show the results of computaions of the spin temperature in the models that we analyze here.
	
	Since the frequency of the 21 cm line $\nu_\mathrm{21}$ is in the Rayleigh-Jeans range of the energy distribution of the CMB, it is convenient to use the brightness temperature $T_\mathrm{br}$ instead of the intensity: $I_\mathrm{\nu}=2k_\mathrm{B}T_\mathrm{br} \nu^2/c^2$. In this case, the useful (eigen) signal is the total intensity in the line, obtained by solving the radiative transfer equation, minus the CMB intensity in the line: $\delta I_\mathrm{\nu}=(I_\mathrm{\nu}-I_\mathrm{\nu} ^\mathrm{CMB})/(1+z)$. The expression for the differential brightness temperature\footnote{We will call it brightness temperature for short.} in the 21 cm line at given redshift $z$ has the form \cite{Madau1997,Zaldarriaga2004,Furlanetto2006,Pritchard2012}:
	\begin{equation}
		T_\mathrm{br}(z)=0.144 x_\mathrm{HI} \frac{\Omega_bh}{\Omega_m^{1/2}} (1+z)^{1/2} \left[1-\frac{T_\mathrm{CMB}}{T_\mathrm{s}}\right]\,\, \mbox{\rm K,}\label{dTbr}
	\end{equation}
	where $x_\mathrm{HI}$, $T_\mathrm{CMB}$ and $T_\mathrm{s}$ are computed for the same redshift $z$. As we can see, the brightness temperature is proportional to the difference between $T_\mathrm{s}$ and $T_\mathrm{CMB}$: if $T_\mathrm{s}<T_\mathrm{CMB}$ then we will have an absorption line, and emission one in the opposit case, $T_\mathrm{s}>T_\mathrm{CMB}$. The line disappears when $T_\mathrm{s}\rightarrow T_\mathrm{CMB}$.
	
	Figures \ref{dTbr_dman}-\ref{dTbr_dmd} show the computation results of the brightness temperature in the 21 cm line of neutral hydrogen from the redshift range $400\le z\le6$, or frequency range $4\le \nu_\mathrm{obs}\le200$ MHz, for models for which the ionization and thermal history has been computed in the previous section. As can be seen, in cosmological models with additional heating (color lines), the dependence of global signal on redshift (tomography) in the 21 cm line of neutral hydrogen differs significantly from the tomography in the standard $\Lambda$CDM model (black lines) and depends on the parameters of the dark matter model.

	\subsubsection{Self-annihilating dark matter}
	\begin{figure}[htb] 
		\includegraphics[width=0.47\textwidth]{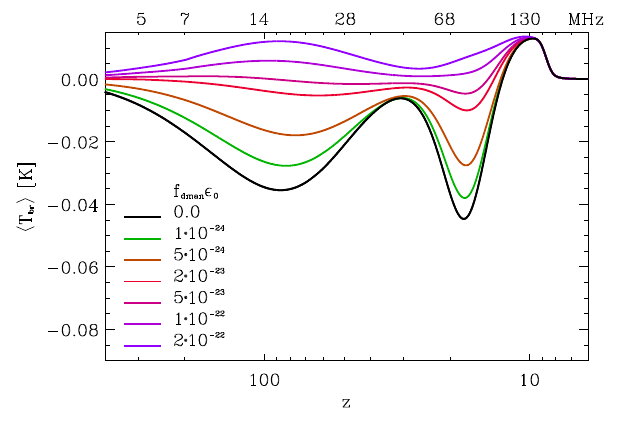}
		\includegraphics[width=0.47\textwidth]{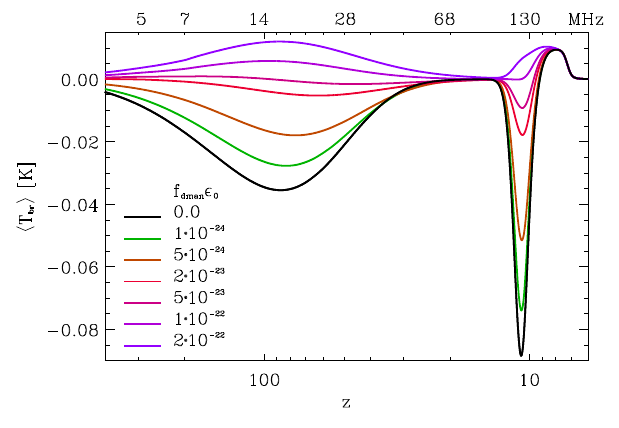}
		\includegraphics[width=0.47\textwidth]{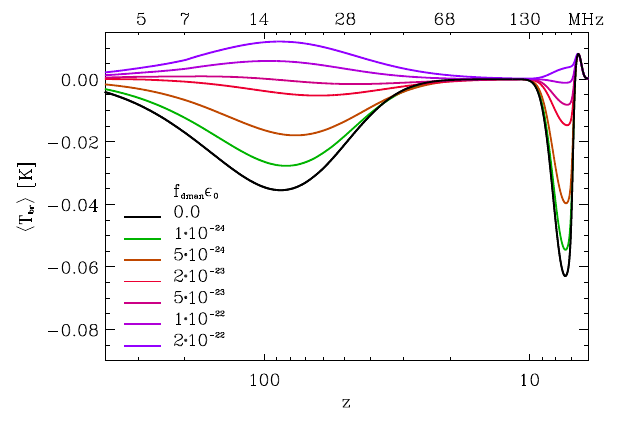}
		\caption{The global signal in redshifted 21 cm line from Dark Ages and Cosmic Dawn in the model with self-annihilating dark matter with different values of the consolidated parameter $f_\mathrm{dman}\epsilon_0$, which increases for lines from bottom to top, and three models of the first light flIIa, flIIb and flIIc (top, middle and bottom panels, accordingly).}
		\label{dTbr_dman}
	\end{figure}
	
	The dependence of global signal in the 21 cm line on $z$ in models with self-annihilating dark matter is sensitive to the values of consolidated parameter $f_\mathrm{dman}\epsilon_0>10^{-25}$ (Fig. \ref{dTbr_dman}). The absorption lines, which are formed during the Dark Ages and Cosmic Dawn, become shallower with increasing the value of consolidated parameter. The Dark Ages absorption line disappears when this parameter become $\sim5\cdot10^{-23}$. The presence of a spectral feature in the Cosmic Dawn signal depends on the parameters of the first light, which forms the position and depth of the absorption line due to the Wouthuysen-Field effect. Further growth of the parameter leads to the appearance of an emission line in the signal from these epochs. However, its amplitude does not exceed 15 mK for permissible values of the consolidated parameter of self-annihilating dark matter.

	\subsubsection{Decaying dark matter}
	
	The differences of the thermal/ionization histories during the Dark Ages in the models with self-annihilating and decaying dark matter particles (Figs. \ref{xT_dman} and Figs. \ref{xT_dmd}) are caused by different dependences of the rates of processes on number densities of dark matter particles: $\propto f_\mathrm{dman}^2n_\mathrm{dm}^2$ in the first model and  $\propto f_\mathrm{dmd}n_\mathrm{dm}$ in the second one. In the Cosmic Dawn epoch the influence of the first sources radiation becomes determinative. It is clearly visible from the comparison of Figs. \ref{xT_dman}-\ref{xT_dmd} with Figs. 5 and 7 in \cite{Novosyadlyj2024}. All these are ``coded`` in the global signal in the 21 cm line from this epochs. Signal from the Dark Ages is sensitive to the consolidated parameter of decaying dark matter $\tau_\mathrm{dmd}/f_\mathrm{dmd}$ when it is $<10^{28}$ s. Signal from the Cosmic Dawn epoch is more sensitive: the depth is noticeably shallower when $\tau_\mathrm{dmd}/f_\mathrm{dmd}<10^{30}$ s.  
	
\begin{figure}[htb] 
		\includegraphics[width=0.47\textwidth]{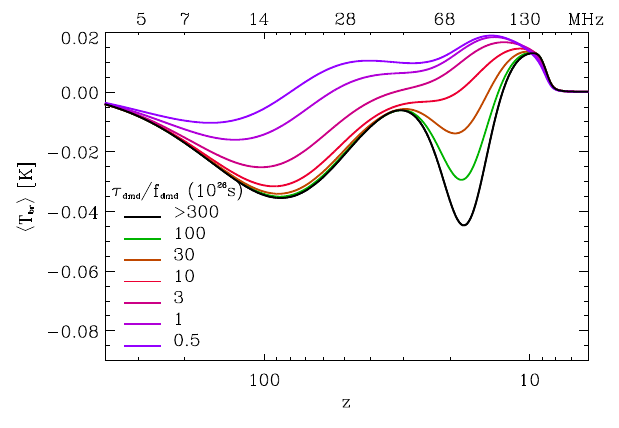}
		\includegraphics[width=0.47\textwidth]{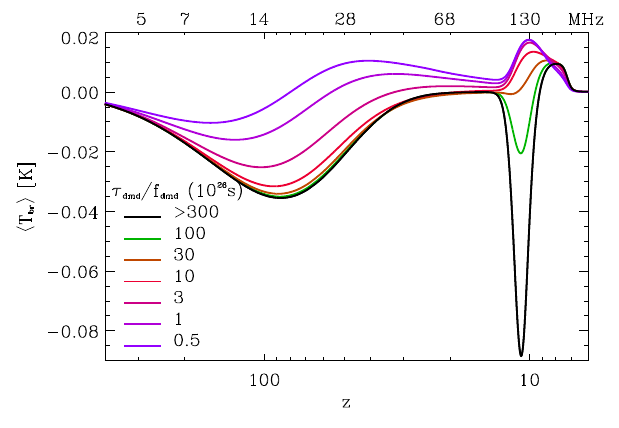}
		\includegraphics[width=0.47\textwidth]{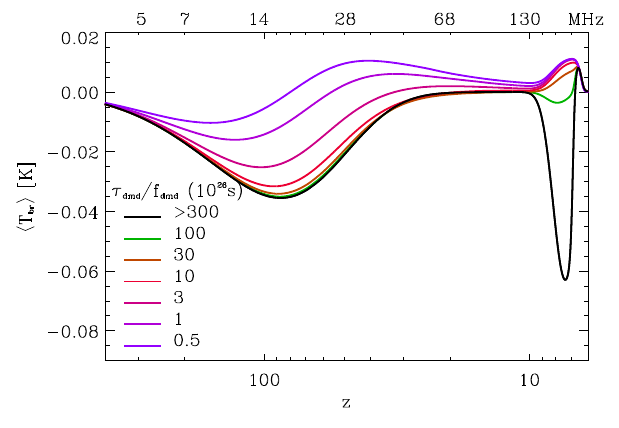}
		\caption{The global signal in redshifted 21-cm line from Dark Ages and Cosmic Dawn in the model with decaying dark matter with different values of the consolidated parameter $\tau_\mathrm{dmd}/f_\mathrm{dmd}$, which decreases for lines from bottom to top, and three models of the first light flIIa, flIIb and flIIc (top, middle and bottom panels, accordingly).}
		\label{dTbr_dmd}
	\end{figure}

	Fig. \ref{dTbr_dmd} shows the dependence of global signal in the redshifted 21 cm line on redshift for models with different values of consolidated parameter $\tau_\mathrm{dmd}/f_\mathrm{dmd}$ in the range $5\cdot10^{25}-10^{28}$ s for three models of the first light, which provide early, middle, and late reionization of hydrogen, as it shown in Fig. 1 and 3. Decreasing of $\tau_\mathrm{dmd}/f_\mathrm{dmd}$ makes the absorption lines less deep and transfers it into wide emission line with amplitude $\lesssim10$ mK at $\sim 20-70$ MHz for values of consolidated parameter close to an upper limit of the considered range. At higher frequencies the spectral feature in the redshifted 21 cm line crucially depends on the model of the first light and evolution of its spectral energy density. Depending on SED the emission hump with amplitude $\sim10-20$ mK may exist in the frequency range $50-150$ MHz. This spectral feature is common for all models of dark matter considered here, stable, self-annihilating and decaying one. A distinguishing feature of the decay model of dark matter from the self-annihilation model is the absorption line in the decameter range at frequencies $\sim7-10$ MHz. 
	
	\section{Discussions and conclusions}
	
	We have estimated the global signal in the redshifted 21 cm hydrogen line formed during the Dark Ages, Cosmic Dawn and Reionization epochs ($6\leq z\leq300$, $5\leq\nu\leq200$ MHz) in the cosmological models with three types of cold dark matter particles (stable, self-annihilating and decaying) and three models of the first light, which provide early, middle, and late reionization of hydrogen (Fig. 1-3). Variations of their parameters change the ionization and thermal histories of the baryonic gas during these epochs, and number density of $Ly_\alpha$ and $Ly_\mathrm{c}$ quanta. The standard $\Lambda$CDM model with Planck parameters \cite{Planck2020,Planck2020a} was taken as the fiducial model. In the papers \cite{Novosyadlyj2023,Novosyadlyj2024} we have analysed separatly the dependences of spectral features of radiation from the Dark Ages on the parameters of self-annihilating and decaying dark matter, and the dependence of the Cosmic Dawn spectral features on the parameters of the model of the first light. Comparison of the obtained results with those results provides a deeper understanding of the relationship between the dependence of the global signal in the 21 cm line on the nature of dark matter particles and the spectral density of radiation of the first light sources. It may be important for the simulating the data of going and future experiments mentioned in the Introduction.
	
	The standard $\Lambda$CDM model with Planck parameters predicts a value of the differential brightness temperature for the signal from the Dark Ages epoch at the center of absorption line of $T_\mathrm{br}\approx-35$ mK at $z\approx87$. The frequency of the line at the absorption maximum for Earth observer is $\approx$16 MHz, and the effective width $\approx$25 MHz. The first light practically does not change the profile of this absorption line even at the upper level of reionization given by Planck. 
	
	In the models with self-annihilating dark matter particles, the global signal in the redshifted 21 cm line is sensitive to the additional ionization and heating when consolidated parameter $f_\mathrm{dman}\epsilon_0>10^{-25}$ (Fig. \ref{dTbr_dman}). As the upper value of this parameter, we took $2\cdot10^{-22}$, at which the optical depth of relic radiation due to Thomson scattering on free electrons is close to the upper $2\sigma$-limit, which follows from the data of \cite{Planck2020}. The value of the fraction of ionized hydrogen for these values of the parameter at $z\sim30$ is in the range of $2\cdot10^{-4}\lesssim x_\mathrm{HII}\lesssim4\cdot10^{-3}$, the gas temperature is in the range of $15\lesssim T_\mathrm{b} \lesssim150$ K (Fig. \ref{xT_dman}). At the lower redshifts, they are determined mainly by the first light model. The spectral feature in the redshifted 21 cm line, formed during the Dark Ages, changes from absorption line with brightness temperature $T_\mathrm{br}\gtrsim-35$ mK up to emission one with $T_\mathrm{br}\lesssim15$ mK. The profile of the spectral feature, formed during Cosmic Dawn epoch, are determined mainly by the evolution of the number density of $Ly_\alpha$ and $Ly_\mathrm{c}$ quanta, but its amplitude is sensitive to the consolidated parameter $f_\mathrm{dman}\epsilon_0>10^{-25}$.
	
	In the models with decaying dark matter particles, the global signal in the redshifted 21 cm line from the Dark Ages ($\nu\sim16$ MHz) is sensitive to the additional ionization and heating when consolidated parameter $\tau_\mathrm{dmd}/f_\mathrm{dmd}\lesssim10^{28}$ s. Signal from the Cosmic Dawn epoch ($\nu\gtrsim70$ MHz) is sensitive when $\tau_\mathrm{dmd}/f_\mathrm{dmd}\lesssim10^{30}$ s (Fig. \ref{dTbr_dmd}). The difference is caused by growth of baryonic gas temperature with time $T_\mathrm{b}\propto (z+1)^{-(1.2\div1.5)}$ at $z_\mathrm{min}\lesssim z<10$, where $z_\mathrm{min}$ corresponds to redshift of the minimum of gas temperature in Fig. \ref{xT_dmd}. It is important that absorption line formed during the Dark Ages does not disapper with decreasing of consolidated parameter, but become less deep and its peak position moves to lower frequency (higher redshift). In the models with lower value of consolidated parameter the emission hump appeares in its short-wavelength wing at $\nu\sim 25-40$ MHz. Its amplitude does not exceed $\sim10$ mK in our models. Its short-wavelength edge is sensitive to the model of the first light. The spectral feature of Cosmic Dawn epoch, absorption line with a narrow emission hump in its short-wavelength wing in the model with stable dark matter particles ($70\lesssim\nu\lesssim170$ MHz), quickly turns into a broad emission hump in the case of decaying particles with lowering their life-time or inreasing their fraction. It disappears in the reionization epoch at $z\sim 8-6$ ($\nu\gtrsim170$ MHz.
	
	The right wing of the emission feature of the Reionization epoch ($z\lesssim8$) is defined crucially by the disappearing of neutral hydrogen atoms via reionization by the first light and slightly depends on the dark matter nature. In contrast, the left wing strongly depends on the first light model and nature of dark matter particles (Figs. \ref{dTbr_dman}-\ref{dTbr_dmd}).
	
	The results presented in Fig. \ref{dTbr_dman}-\ref{dTbr_dmd} show that the degeneracy of the dependence of the amplitude of the global signal in the 21 cm line from the Cosmic Dawn epoch ($50\le\nu\le100$ MHz) on the parameters of the first light model as well as the dark matter self-annihilation/decay models can be eliminated or significantly weakened by the data on the global signal from the Dark Ages ($10\le\nu\le50$ MHz). It is possible to distinguish the model of self-annihilation of dark matter particles from the model of their decay using precision tomography in a wide frequency range ($10\le\nu\le100$ MHz) based on the known evolution of the number densities of $Ly_\alpha$ and $Ly_\mathrm{c}$ quanta. However, this will be a challenge even for the next generation of radio telescopes of this frequency range.
	
	Obtained results are important for simulating of possibilities of extracting cosmological information from going and planned groundbased, space and Moon far-side experimens. They also prove that wide frequency detection of the global signal in the redshifted hydrogen 21 cm line from the Dark Ages and Cosmic Dawn epochs can be powerfull cosmological test of dark matter nature and scenarios of the formation of the first light sources. 
	
	\section*{Acknowledgements}
	This work is done in the framework of the project ``Tomography of the Dark Ages and Cosmic Dawn in the lines of hydrogen and the first molecules as a test of cosmological models'' (state registration number 0124U004029) supported by National Research Foundation of Ukraine. BN thanks to International Centre of Future Science of Jilin University for hospitality. The authors are grateful to Pavlo Kopach for technical assistance in the preparation of the paper. 
	
	%\section*{Data availability}
	%The data underlying this article will be shared on reasonable request to the corresponding author Bohdan Novosyadlyj (bnovos@gmail.com).

\end{document}